\documentclass[conference]{IEEEtran}
\IEEEoverridecommandlockouts
\usepackage{cite}
\usepackage{amsmath,amssymb,amsfonts}
\usepackage{algorithmic}
\usepackage{graphicx}
\usepackage{textcomp}
\usepackage{xcolor}
\usepackage{subcaption}
\usepackage{hyperref}
\hypersetup{
	colorlinks=true,
	linkcolor=blue,
	filecolor=magenta,      
	urlcolor=blue,
}

\def\BibTeX{{\rm B\kern-.05em{\sc i\kern-.025em b}\kern-.08em
    T\kern-.1667em\lower.7ex\hbox{E}\kern-.125emX}}
\begin{document}

\title{Trust from Ethical Point of View: Exploring Dynamics Through Multiagent-Driven Cognitive Modeling\\
}

\author{\IEEEauthorblockN{ Abbas Tariverdi}
	\IEEEauthorblockA{
		\textit{Department of Physics} \\
		\textit{University of Oslo}\\
		Oslo, Norway \\
		abbast@uio.no}
}

\maketitle

\begin{abstract}
The paper begins by exploring the rationality of ethical trust as a foundational concept. This involves distinguishing between trust and trustworthiness and delving into scenarios where trust is both rational and moral. It lays the groundwork for understanding the complexities of trust dynamics in decision-making scenarios. Following this theoretical groundwork, we introduce an agent-based simulation framework that investigates these dynamics of ethical trust, specifically in the context of a disaster response scenario. These agents, utilizing emotional models like Plutchik's Wheel of Emotions and memory learning mechanisms, are tasked with allocating limited resources in disaster-affected areas. The model, which embodies the principles discussed in the first section, integrates cognitive load management, Big Five personality traits, and structured interactions within networked or hierarchical settings. It also includes feedback loops and simulates external events to evaluate their impact on the formation and evolution of trust among agents. Through our simulations, we demonstrate the intricate interplay of cognitive, emotional, and social factors in ethical decision-making. These insights shed light on the behaviors and resilience of trust networks in crisis situations, emphasizing the role of rational and moral considerations in the development of trust among autonomous agents. This study contributes to the field by offering an understanding of trust dynamics in socio-technical systems and by providing a robust, adaptable framework capable of addressing ethical dilemmas in disaster response and beyond. The implementation of the algorithms presented in this paper is available at this GitHub repository: \url{https://github.com/abbas-tari/ethical-trust-cognitive-modeling}.
\end{abstract}


\section{Introduction}
The German-American philosopher Max Otto (1876–1968) made a significant statement about trust: "The deepest source of a man’s philosophy, the one that shapes and nourishes it, is faith or lack of faith in mankind" \cite{otto1940human}. Otto suggests that our philosophy is influenced by our level of faith or trust in others. This trust shapes our worldview and understanding of humanity.

First we delves into the critical distinction between trust and trustworthiness, highlighting that while individuals have a moral duty to be trustworthy, exhibiting qualities like honesty and fairness, there isn't a similar moral obligation to trust others. This section explores the nuanced relationship between these concepts, guided by philosophical literature. Works such as those by Simion and Willard-Kyle \cite{simion2023trust}, D’Arms and Kerr \cite{d2015trust}, and Faulkner \cite{faulkner2014moral} discuss the moral imperatives surrounding trustworthiness. Hills \cite{hills2023trustworthiness} challenges the notion of trustworthiness as mere reliability, positioning it as a virtue. These perspectives collectively argue for the moral significance of embodying trustworthiness, a key aspect that facilitates positive social interactions and reciprocity. However, the act of trusting remains a nuanced, situation-dependent decision, not mandated by moral duty.

Furthermore we elaborate on the concept of 'Trust Limits,' emphasizing the multidimensional nature of trust, which includes self-trust and trust in external entities such as individuals, objects, and institutions. Each aspect of trust, whether it be honesty, competence, or modesty, possesses its own trustworthiness spectrum, necessitating discernment in what and to what extent trust is vested. This discussion leads to an exploration of 'General Trust,' which examines trust based on the comparison of benefits and drawbacks, as suggested by Barbalet \cite{barbalet2009characterization} and Bormann \cite{bormann2021trust}. It also considers the three elements of trust: the trustor, the trustee, and the bounds of trust, and how some individuals may exhibit generalized trust, potentially leading to negative consequences. Empirical studies in various contexts, such as home-school relations and international alliances, have investigated these dynamics, highlighting the importance of consistency in behavior for trust development. The section concludes with an analysis of 'Trust Rationality,' outlining scenarios where trust can be rational, and touches on the relationship between morality and rationality in trust, drawing insights from different philosophical viewpoints like those of Hobbes, Kant, Sidgwick, and Hume.

The transitions from the theoretical underpinnings of trust to its computational modeling, it acknowledges the complex interplay of moral, psychological, and rational aspects in human relationships. Trust necessitates a balance between the trustor and trustee within specific bounds, highlighting the need for discerning and situational trust. Computational models offer a novel perspective in understanding these dynamics. Recent advancements, as pioneered by Marsh (1994) \cite{marsh1994formalising}, have led to models that capture the probabilistic nature of human behavior and trust dynamics using mathematical techniques and algorithmic approaches. Examples include models employing a partially observable Markov decision process to predict human trust dynamics \cite{williams2023computational} and those predicting trust levels based on nonverbal cues in social interactions \cite{lee2013computationally,jayagopi2011computational}. While these models may not fully encapsulate the spectrum of human trust, they provide valuable insights, particularly in policy formulation and technology design. The following sections delve into agent-based modeling, illustrating how computational simulations can model ethical trust in a controlled environment, thereby offering a microcosm for studying the evolution and dynamics of trust.

As the contribution of this study, we present an Agent-based Cognitive Computational Model, designed to simulate ethical trust in collective decision-making processes. Central to our model are Emotional Agents, equipped with a rich array of psychological and social attributes, enabling the simulation of human-like behaviors and interactions. These agents employ Plutchik's Wheel of Emotions for a complex emotional model influencing decision-making, mirroring human cognitive processes in trust and ethics. They also exhibit memory and learning capabilities, reflecting on past interactions to evolve future behaviors, a nod to the impact of experience in ethical and trust dynamics. Our model incorporates elements like social influence, cognitive load, personality traits, and variable interactions, to mirror the multifaceted nature of human decision-making. With feedback loops and external events, it captures the dynamic and unpredictable nature of trust development. This model is a modest contribution towards understanding the nuanced interplay of trust and ethics in social systems, offering insights valuable for research in social dynamics, autonomous system design, and ethically aware AI, thereby marrying computational methodology with psychological and sociological theory to shed light on collective ethical behavior.

 This paper delves into the intricate relationship between rationality and trust, structured as follows: Section II, 'A Foundational Concept: Rationality of Ethical Trust,' establishes the fundamental concepts and theoretical underpinnings of rational trust, setting the stage for further exploration. Section III, 'From Theoretical Foundations to Computational Models,' transitions from theory to practical application, showcasing how these concepts can be modeled computationally. In Section IV, 'Agent-based Cognitive Modeling of Ethical Trust,' we present a cognitive approach to modeling ethical trust through agent-based simulations. Section V, 'Case Study: Disaster Response Decision-Making,' applies these models in a real-world context, specifically in the domain of disaster response decision-making. Finally, Section VI serves as the 'Conclusion,' where we encapsulate our findings and underscore the importance of understanding trust's rationality, especially in scenarios beyond the scope of this study, where trust, while psychologically and subjectively beneficial, may lack objective rationality.

\section{a foundational concept: rationality of ethical trust}
\subsection{Trust and Trustiness}

It's crucial to distinguish between trust and trustworthiness. While individuals have a moral obligation to embody trustworthiness through honesty, modesty, fairness, and promise-keeping, there isn't a moral mandate to extend trust to others. In essence, one's moral duty lies in living a life that warrants trust from others, thereby fulfilling the obligation to be trustworthy. While individuals may not be morally obligated to trust others, there's a moral imperative to be trustable - embodying qualities like honesty, fairness, and reliability that engender trust from others. Let's delve into why living in a trustable way is seen as a moral obligation according to some recent philosophical literature.  The work by Mona Simion and Christopher Willard-Kyle \cite{simion2023trust} possibly discusses the moral imperatives surrounding trustworthiness. In \cite{d2015trust},  the authors explore trustworthiness in light of situationism, hinting at the moral implications of being consistent in one's trustworthy behavior. The work \cite{faulkner2014moral} on the moral obligations of trust, discusses the second-personal nature of moral obligation, which aligns with the interpersonal character of trustworthiness. This work hints at the reason for the belief supplied by testimony supporting trust, which might be seen as a reflection of one's trustworthiness. The discussion on trustworthiness, responsibility, and virtue, \cite{hills2023trustworthiness} challenges the notion that trustworthiness is merely a kind of reliability, arguing instead for its recognition as a virtue. This aligns with the moral obligation to embody virtues in one's conduct.

Are humans morally obligated to trust? Evidently not. Unlike the moral duty to be trustworthy, extending trust is not a moral obligation. Therefore, it's vital to differentiate trust from trustworthiness and ascertain the scenarios where moral trust is warranted. This discussion focuses on trust, not trustworthiness, aligning with the literature that often differentiates between the moral duty to be trustworthy and the act of placing trust in others. 

The asymmetry between the moral obligations of trust and trustworthiness arises from the inherent virtues of trustworthiness, reflecting an individual's moral integrity and promoting societal cooperation \cite{simion2023trust, hills2023trustworthiness}. Trustworthiness, tied to moral virtues, fosters positive social interactions and reciprocity, while the act of trusting is a nuanced decision contingent on various factors, not mandated by moral duty \cite{d2015trust,faulkner2014moral}.

\subsection{Trust Attendant}

The discourse on trust delineates a nuanced dichotomy between self-trust and trust in others or entities. Self-trust, posited as a moral necessity, underscores an individual's duty to trust one's own capacities and judgments intermittently, anchoring the foundation for interpersonal trust dynamics \cite{zalta1995stanford}. This notion complements the previously discussed asymmetry in moral obligations concerning trust and trustworthiness, where individuals are morally obligated to embody trustworthiness, fostering a conducive milieu for social interactions and reciprocity, while the act of trusting others remains a contextual, discretionary decision \cite{simion2023trust,d2015trust,hills2023trustworthiness}. The complexity of trust extends to interactions with individuals, objects, and institutions, each bearing distinct considerations that enrich the philosophical understanding of trust's multifaceted nature. The moral landscape surrounding trust encapsulates a spectrum of obligations and decisions, influenced by the interplay of self-trust and external trust, underscoring the context-dependent intricacies inherent in trust relationships.

\subsection{Trust Limits}

Trust, inherently multi-dimensional, oscillates between self and external entities—individuals, objects, or institutions, echoing the philosophical discourse that dissects trust into various components or attributes that can be individually trusted or distrusted \cite{carter2020ethics}. The crux, however, resides in discerning the facets of trust—be it honesty, competence, or modesty, each with its own spectrum of trustworthiness. This nuanced landscape of trust propels the discourse into 'Trust Limits', underscoring the necessity to delineate what and to what extent trust is vested, whether in oneself or others. For instance, one's trust in another's honesty may not translate to trust in their generosity, illuminating the segmented nature of trust. Such trust limits, pivotal in navigating interpersonal and intrapersonal interactions, beckon a deeper exploration into the ethics of trust. The spectrum of trustworthiness, spanning from complete distrust to blind trust, embodies the essence of trust limits, urging a discernment in what and whom we trust, and to what extent. This discourse not only enriches the understanding of trust but also underscores the importance of discernment and awareness in fostering a culture of trust imbued with respect, recognition, and moral integrity, resonating with the philosophical narrative that explores the ethics and epistemology of trust  \cite{carter2020ethics}.

\subsection{General Trust}
The necessity of trust in a given relationship or interaction may be evaluated based on the comparison of benefits and drawbacks \cite{barbalet2009characterization}. Trust is deemed necessary when the anticipated or realized benefits outweigh the potential or actual drawbacks, promoting a more favorable outcome for the trusting party \cite{bormann2021trust,evans2018reputational} This perspective emphasizes the practical outcomes of trust, aligning with the consequentialist principle of maximizing positive consequences while minimizing negative ones \cite{miklos2019consequentialism}.

Trust comprises three elements: 1) the trustor, 2) the trustee, and 3) the bounds of trust. Some individuals, due to their upbringing or inherent traits, may prioritize the first element over the latter two. They may exhibit generalized trust, not distinguishing between different trustees and not considering the bounds of trust. For instance, when they trust someone in one aspect, they may extend trust to other unrelated aspects, overlooking the specificity of trust.

Trust serves as a cornerstone in societal interactions \cite{lewis2012social}, underpinning crucial outcomes such as reciprocity, collective action, and social order \cite{schilke2021trust}. Sociological examinations often delve into trust within a broader social landscape, emphasizing an individual’s willingness to embrace vulnerability or engage in risky interactions \cite{simon2020routledge}.

A variety of empirical studies have explored trust in different contexts. For instance, in home-school relations, trust has been extensively studied, with 79 peer-reviewed quantitative empirical studies analyzed over two decades
\cite{shayo2021conceptualization}. Similarly, in international alliances, trust is dissected into weak, semi-strong, and strong forms, with empirical evidence showing the importance of consistency in behavior for trust development \cite{parkhe1998understanding}.

From a social psychology perspective, disregarding the second and third facets of trust—either displaying indiscriminate trust (toward individuals, objects, or institutions) or entrusting someone (or an object or institution) in all matters—can be perilous. Hence, it's advised against nurturing children in such a manner. This undiscriminating trust can stem from two sources:
\begin{itemize}
	\item Genetics: Some individuals may have a genetic inclination toward a broad-spectrum trust, known as Dispositional Trust, where trust is extended in a generalized manner across various entities and situations \cite{evans2018reputational,patent2022dysfunctional}.
	\item Upbringing: On the other hand, some are not genetically predisposed but are nurtured to trust indiscriminately, termed as Learned General Trust \cite{ayoub2021modeling}.
\end{itemize}

Regardless of the origin—be it genetic predisposition or upbringing—the outcome, as per social psychology, is unfavorable. Unrestrained trust is detrimental primarily because all humans are not trustworthy. Secondly, trust should be situational and rational, not universal. Generalized trust can lead to negative judgments, especially when individuals do not discern between trustworthy and untrustworthy entities \cite{evans2018reputational}. Moreover, such trust behavior can adversely affect one's happiness and mental quality of life \cite{lin2021mediated}. Experiences of loss of social trust are often recounted, further emphasizing the need for a more discerning approach to trust \cite{doyle_fragile_2023}.

\subsection{Trust Rationality}\label{trust_rationality}

Trust rationality pertains to the logical foundation underpinning trust in various scenarios. It's essential to discern that rationality in this context refers to both understandability and defensibility of trust—i.e., the logical reasons for placing trust and the potential benefits thereof. This section explores five scenarios where trust can be deemed rational based on mutual benefits, shared objectives, or established moral grounds. However, generalized trust beyond these scenarios may lead to unwarranted risks or misplaced expectations.

\begin{itemize}
	\item Trust in Affectionate Relationships: Trusting individuals who exhibit affection or friendship towards you is rational as mutual benefits are inherent in such relationships. The trust here relies on the predictability and dependability of actions that promote shared interests \cite{rempel1985trust}.
	
	\item Trust in Shared Benefits: When mutual benefits are apparent—as in the case of co-passengers on an airplane aiming for a safe journey—trust is rational. Here, trust is specific to the shared objective and doesn't extend beyond that scenario.
	
	\item Trust in Moral Integrity: Trusting individuals displaying a minimum level of moral integrity is rational. However, defining 'minimum morality' requires a clear set of ethical benchmarks. Trust in this scenario relies on the perceived moral alignment and the potential for mutual respect and understanding \cite{rathbun2009takes}.
	
	\item Monitoring-induced Trust:
	Trust becomes rational when there's a mechanism to monitor and ensure the fulfillment of responsibilities, as seen in professional settings. This form of trust, often termed Monitoring-induced Trust, is conditional on the oversight ensuring the trustee's accountability \cite{yang2022towards,francke2022trust}.
	
	\item Trust Based on Historical Reliability:
	
	Placing trust in individuals who have historically upheld trust can be rational but with a lower degree of certainty. This scenario acknowledges the possibility of betrayal despite a history of trustworthiness, thus making it a less robust form of rational trust.

\end{itemize}

	These five scenarios illustrate different facets of trust rationality, highlighting the importance of situational analysis, mutual benefits, and moral alignment in rationalizing trust. Generalized trust, devoid of these considerations, may lead to undesired consequences, underscoring the need for a discerning approach to trust.
 
	\subsection{Morality and Rationality}

The relationship between morality and rationality can be dissected through various philosophical lenses, each offering a unique perspective on how moral decisions align or conflict with rational thinking. This section draws from the discourse presented in the referenced work \cite{baier1982conceptual}, aiming to elucidate the correlation between morality and rationality through four primary philosophical views, supplemented with interpretations to enhance comprehension:

\begin{itemize}
	\item Hobbesian View:

	Stemming from Hobbes, this perspective posits that moral or immoral actions are deemed so based on their rational or irrational nature, respectively, in a given set of conditions. Essentially, actions aligned with one's interests or desires under certain circumstances are viewed as moral and rational, while those contrary are seen as immoral and irrational \cite{london1998virtue,meyers2013hobbes}.

	\item Kantian View: Drawing from Kant, this stance concurs with Hobbesians that immorality clashes with reason but diverges by asserting that moral precepts can sometimes conflict with self-interest. This view underscores the potential tension between adhering to moral principles and pursuing personal interests \cite{uleman2010introduction,wood2007kantian}.
	
	\item Sidgwickian View: Originating from Sidgwick, this perspective holds that while it's invariably rational to be moral, there are instances where immorality may also be rational. This view acknowledges a nuanced interplay between moral and rational judgments, allowing for a broader spectrum of rational responses to moral dilemmas \cite{hooker2000sidgwick,brink1992sidgwick,phillips2022sidgwick,schneewind1977sidgwick}.
	
	\item Humean View: Rooted in Hume's philosophy, this position contends that morality and rationality are not necessarily intertwined. This view challenges the inherent alignment or discord between moral and rational judgments, promoting a more flexible understanding of their relationship\cite{cohon2004hume,botros2006hume}.

\end{itemize}

	Each of these philosophical viewpoints offers a framework to explore the intricacies between morality and rationality, providing a rich tapestry of thought to understand the manifold ways in which moral judgments interact with rational deliberation. Through this nuanced examination, we can appreciate the complexity of moral and rational interplay in decision-making and ethical considerations.

Prior discussions spotlight a discerning approach to trust, underscoring the imprudence of indiscriminate trust. It is also highlighted that trust granted in one dimension does not obligate trust in all dimensions concerning an entity or individual. Five scenarios of rational trust were outlined, showcasing a gradation of rationality in the scenarios discussed in \ref{trust_rationality} . A concern emerges about the completeness of this categorization; without validating the exhaustive nature of this list, the assertion that only these five scenarios epitomize rational trust may be challenged.

Acknowledgment of the rational foundation of trust in the identified scenarios, coupled with an endorsement of a close nexus between rationality and morality, leads to the deduction that trust, both rational and moral, is encapsulated solely within these demarcated scenarios. Conversely, trust beyond these scenarios is seen as devoid of both rational and moral grounding, adhering to the notion that rationality should be in harmony with morality.

Addressing the concern, the list, albeit possibly non-exhaustive, offers a substantial framework to navigate the interplay of rationality, morality, and trust. The articulation of these cases is rooted in practical and philosophical considerations, illuminating common scenarios where trust rationality intersects with morality. This list acts as a springboard for deeper exploration, open to extension or refinement through further empirical or philosophical inquiry. Thus, the potential non-exhaustiveness of the list does not significantly undermine the stance but instead lays groundwork for enriched discourse and examination in the field of trust rationality and morality.

The synchronization between rationality and morality within the trust framework is accentuated here. Yet, this discourse could be fortified by further empirical or philosophical scrutiny to robustly substantiate the claims presented.

This narrative accentuates the intertwined relationship between rationality and morality within the trust framework. However, a more thorough exploration and justification of the specified cases of rational trust, potentially through empirical or philosophical examination, could bolster the defense of the claim that these scenarios exclusively represent rational and moral trust.

\section{From Theoretical Foundations to Computational Models}

As the aforementioned sections have explored, trust plays an intricate role in human relationships, involving a delicate interplay of moral, psychological, and rational considerations. At its core, trust requires a delicate balance between the trustor and trustee, taking into account the specific bounds within which trust operates. Through a comprehensive understanding of these dynamics, we've highlighted the critical need for discernment in trust, arguing against the pitfalls of generalized trust and instead championing trust that is situational, rational, and, to an extent, moral.

However, while these insights provide a robust theoretical foundation and as we transition into an age of interconnectedness where algorithms, artificial intelligence, and big data influence numerous aspects of our lives, it becomes imperative to study trust through computational lenses as well.

The immediate question that arises is: How can the multifaceted nature of trust, rooted in human emotions and moral complexities, be distilled into computational models? Such models, while inherently simplified, can offer invaluable insights into the dynamics of trust on a larger scale, potentially assisting in policy formulation, technology design, and more.

As we embark on this exploration, it's essential to note that while computational models might not capture the entirety of human trust dynamics, they serve as a potent tool for simulating and analyzing specific components of trust, such as trustworthiness, reliability, and reputation, in controlled environments. This controlled analysis can unearth patterns and insights that might be difficult, if not impossible, to discern in the real world due to the multitude of confounding variables.

Agent-based modeling provides one such avenue, bringing the theoretical constructs of trust into the realm of computational simulations. By establishing autonomous agents that operate based on defined rules and behaviors, these simulations can offer a microcosm of society, allowing for a closer examination of how trust develops, evolves, and potentially erodes in different scenarios.

In the subsequent sections, we delve deeper into the mechanics of agent-based modeling for trust, elucidating how ethical trust can be both modeled and understood in a simulation framework. 

\section{Agent-based Cognitive Modeling of Ethical Trust}

The proposed Agent-based Cognitive Computational Model represents a pioneering approach to the simulation of ethical trust in collective decision-making processes. Central to the model are Emotional Agents, sophisticated entities that exhibit a breadth of psychological and social characteristics, allowing for an intricate simulation of human-like behavior and interaction.

\begin{itemize}
    \item  Complex Emotional Models: The agents are equipped with an emotional spectrum derived from Plutchik's Wheel of Emotions, enabling a multifaceted influence of emotional states on decision-making. This complexity allows the agents to demonstrate behaviors that align with contemporary understandings of the role emotions play in human cognitive processes, particularly in the context of trust and ethical decision-making.

    \item Memory and Learning: Agents are capable of recalling past interactions and adjusting future behaviors, encapsulating the human ability to evolve based on historical data. This feature underscores the role of experience in shaping ethical stances and trustworthiness, a fundamental aspect of cognitive development and social interaction.

    \item Social Influence: The model incorporates the concept of social conformity, where agents may align with peer behaviors. This element reflects the social fabric's impact on individual decision-making, acknowledging the cognitive biases inherent in human psychology.

    \item Cognitive Load: By introducing a cognitive load factor, the model accounts for the limitations of processing capacity, akin to human decision-making under stress or information overload. This inclusion is crucial in modeling the reliance on heuristic and emotional cues in complex ethical decisions.

    \item Personality Traits: The integration of the Big Five personality traits allows for diverse interaction patterns and learning behaviors among agents. This diversity acknowledges the deep connection between personality and ethical decision-making, offering a nuanced view of trust dynamics.

    \item Variable Interactions: The model simulates realistic social networks through preferential interactions among agents, reflecting the natural human inclination towards forming selective bonds and communities.

    \item Feedback Loops: Feedback mechanisms within the model enable trust to evolve dynamically, resonating with the nonlinear nature of trust development in human relationships. Such feedback loops provide insight into the reinforcement or degradation of trust over time.

    \item External Events: The inclusion of external events introduces an element of unpredictability and environmental interaction, vital for understanding the volatility of trust and consensus in the face of real-world events.

    \item Diversity of Opinions: Starting with a broad spectrum of initial opinions, the model recognizes the complexity of individual perspectives, crucial for simulating authentic group dynamics and decision-making pathways.
\end{itemize}

The cognitive processes of the agents involve various strategies to adapt opinions and trust levels, mimicking different ethical decision-making approaches observed in humans. The strategies range from consensus-seeking behaviors, like averaging peer opinions, to more individualistic approaches, such as contrarian reactions to majority views.

By iterating the model over extensive simulations, it becomes possible to observe the long-term evolution of consensus and ethical trust among agents. This capability is especially relevant for applications in social dynamics research, autonomous system design, and the development of ethically aware AI. The model is a testament to the potential of computational simulations to capture and explore the intricate web of factors that influence trust and ethical decision-making within social systems.

In conclusion, this Agent-based Cognitive Computational Model bridges computational methodology with deep psychological and sociological theory to illuminate the underpinnings of collective ethical behavior.

\section{Case Study: Disaster Response Decision-Making}
To illustrate the performance of the Agent-based Cognitive Computational Model, let’s define a scenario that encompasses the various facets of the model, demonstrating how the agents interact, adapt, and evolve their trust and opinions over time.

In a disaster response scenario, a group of autonomous agents represents different humanitarian organizations tasked with deciding on the allocation of limited resources to various affected areas. Each organization has a distinct priority, opinion on the urgency of needs, and approach to the disaster, reflecting real-world diversity.

This scenario not only demonstrates the model's ability to simulate complex, real-world decision-making but also showcases the rich, emergent behaviors that arise from the interplay of cognitive, emotional, and social factors within an ethical trust framework. The outputs provide a nuanced understanding of how autonomous agents might navigate disaster response, a task with profound ethical implications and the need for trustful cooperation.

\begin{figure}[htbp]
\centerline{\includegraphics[width=\columnwidth]{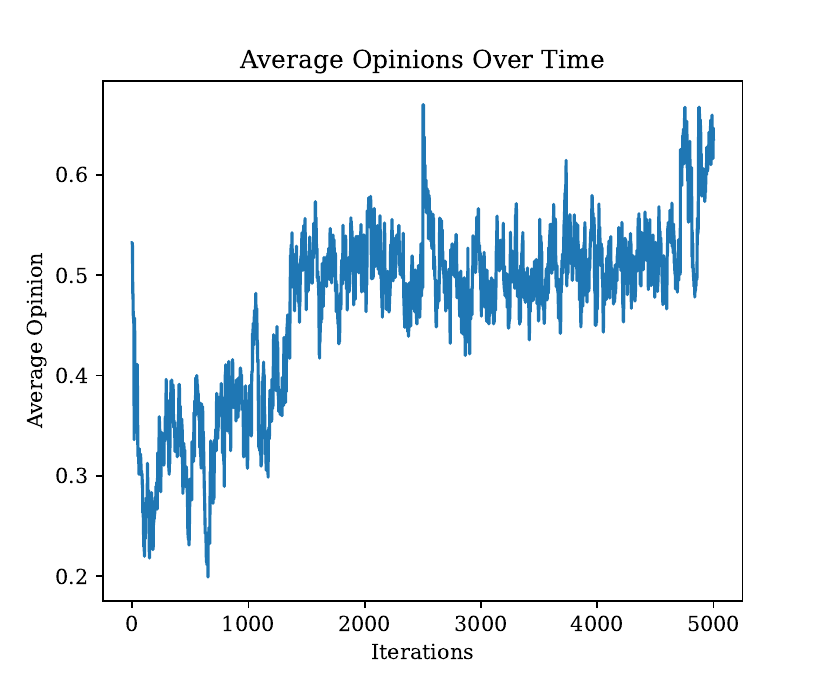}}
\caption{Average Opinions Over Time in a Disaster Response Scenario.}
\label{opinions_over_time}
\end{figure}

As shown in Figure~\ref{opinions_over_time}, the decision-making process in a disaster response scenario is characterized by fluctuating consensus levels among autonomous agents. These agents, representing various humanitarian organizations, are tasked with allocating limited resources to affected areas. The plot illustrates the average opinions over time, capturing the dynamic negotiation and agreement process across 5000 iterations. The variations highlight the complexity of achieving a unified strategy in the face of divergent priorities and the evolving nature of the disaster.

\begin{figure}[htbp]
\centerline{\includegraphics[width=\columnwidth]{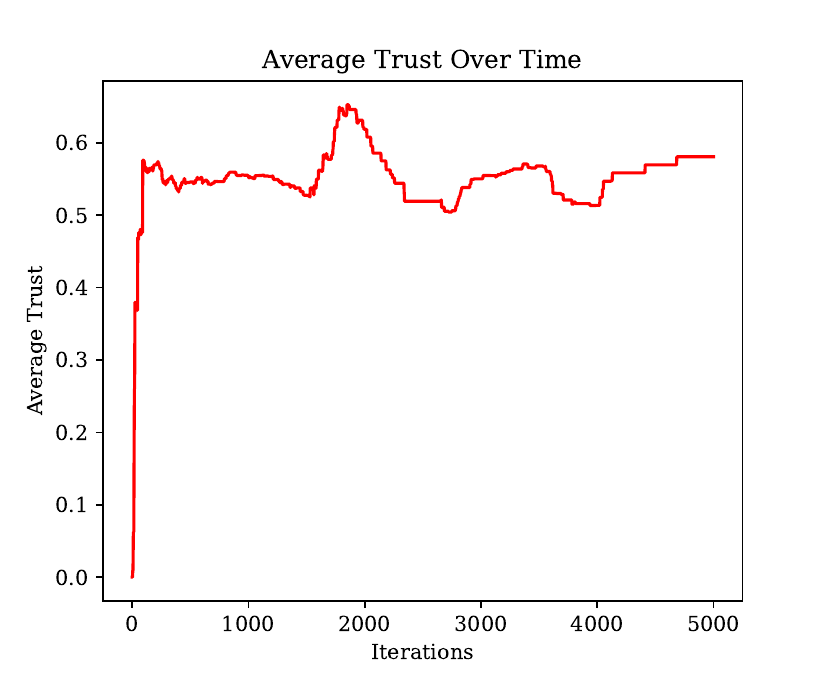}}
\caption{Average Trust Over Time in a Disaster Response Scenario.}
\label{fig:avg_trust_over_time}
\end{figure}

Figure~\ref{fig:avg_trust_over_time} depicts the trend in average trust among agents over time. The red line indicates how trust fluctuates as agents interact and update their opinions and trust levels. Over the course of 5000 iterations, this graph shows the evolution of trust, which is critical in coordinating efforts and sharing resources among the autonomous agents representing different humanitarian organizations in a disaster response scenario. It is evident from the trend that trust dynamics are volatile, reflecting the ongoing reassessment of strategies and partnerships.

\begin{figure}[htbp]
\centerline{\includegraphics[width=\columnwidth]{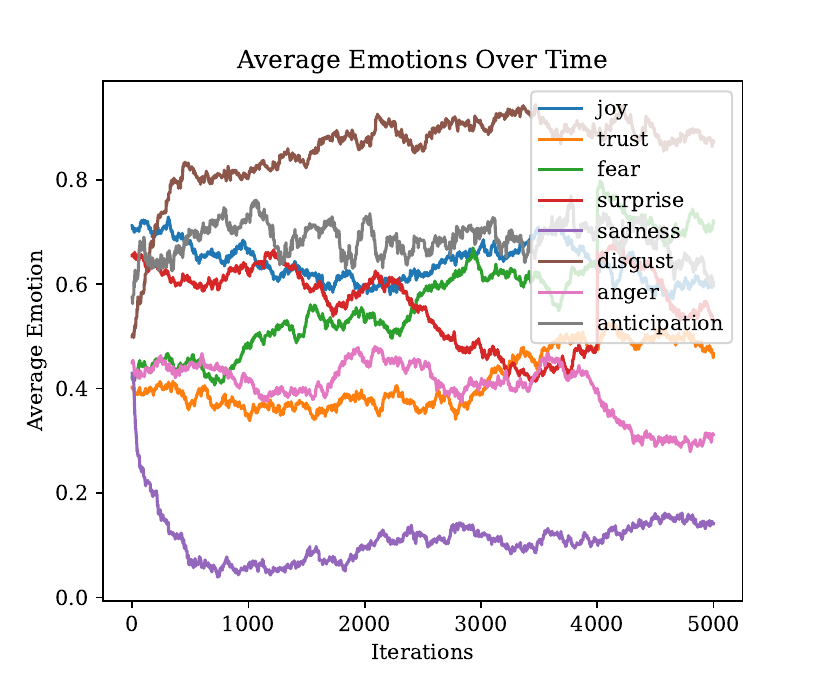}}
\caption{Dynamics of Average Emotions Over Time Among Agents, including an emotional representation of trust.}
\label{fig:emotions_over_time}
\end{figure}

The complex interplay of emotions among autonomous agents in a disaster response scenario is visualized in Figure~\ref{fig:emotions_over_time}. Each line represents the average intensity of a particular emotion over 5000 iterations, such as joy, trust, fear, surprise, sadness, disgust, anger, and anticipation. Notably, the "trust" emotion depicted here differs from the "Average Trust" shown in a separate figure. While the former captures the fluctuating sentiment of confidence among agents, the latter quantifies the reliability based on the history of interactions. These emotional dynamics are crucial for understanding the agents' behavior as they influence decision-making processes and resource allocation strategies. The variability across different emotions underscores the adaptive emotional responses to the changing circumstances of the disaster environment, while the "Average Trust" metric informs the strategic level of collaboration.

\begin{figure}[htbp]
\centerline{\includegraphics[width=\columnwidth]{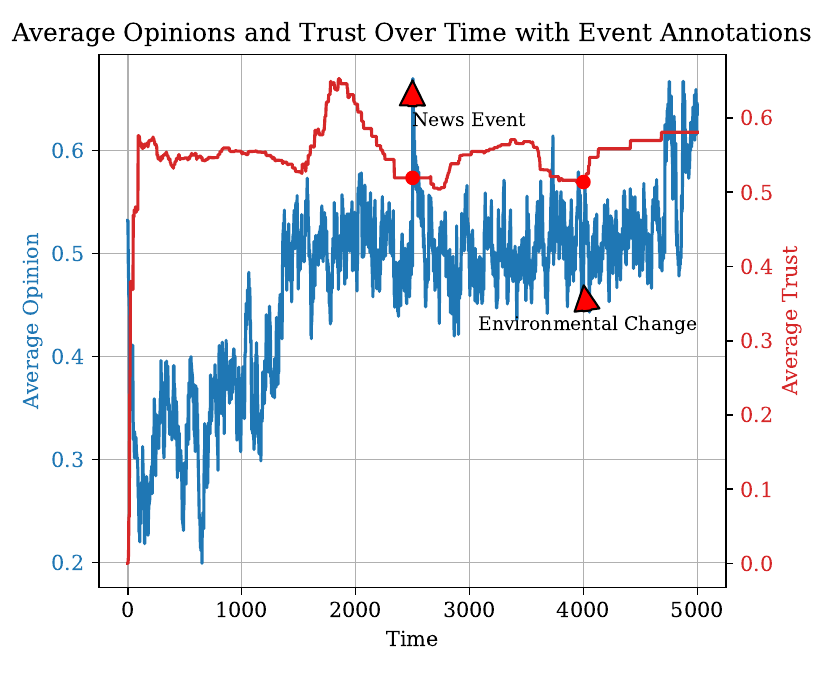}}
\caption{Interplay of Average Opinions and Trust Over Time with Key Event Annotations.}
\label{fig:opinionstrustevents}
\end{figure}

Figure~\ref{fig:opinionstrustevents} provides a dual-axis visualization of the evolution of average opinions and trust among autonomous agents over time, annotated with significant events that influenced these metrics. The blue line represents the average opinion, while the red line indicates the average trust. Specific time points are marked to denote external events, such as a 'News Event' at iteration 2500 and an 'Environmental Change' at iteration 4000, which had notable impacts on the agents' opinions and trust levels. These annotations serve to highlight the correlation between external stimuli and internal agent states, offering insights into the agents' adaptive mechanisms in response to new information and environmental shifts.

\begin{figure}[htbp]
\centering
\begin{subfigure}{.5\textwidth}
  \centering
  \includegraphics[width=\linewidth]{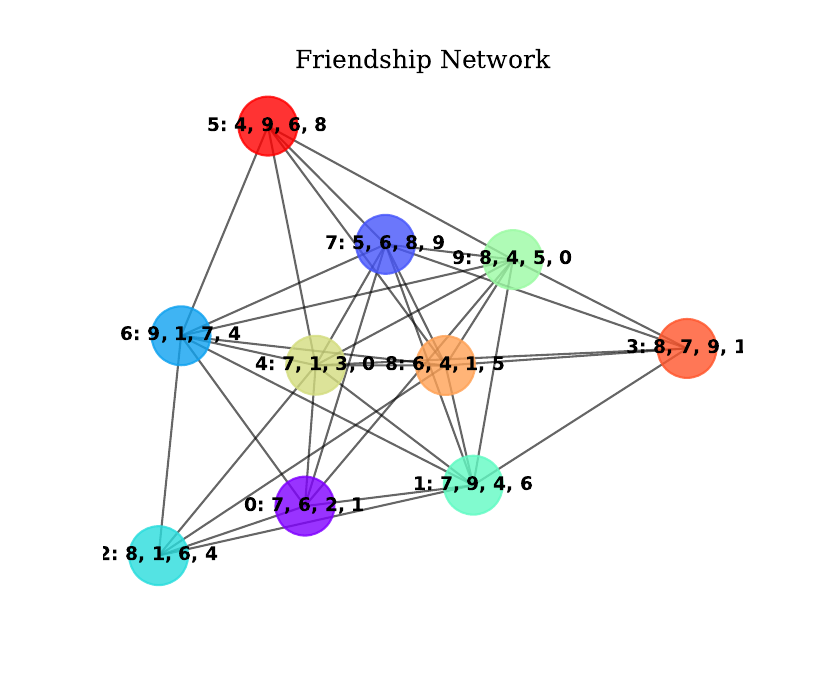}
  \caption{Friendship Network}
  \label{fig:sub1}
\end{subfigure}

\begin{subfigure}{.5\textwidth}
  \centering
  \includegraphics[width=\linewidth]{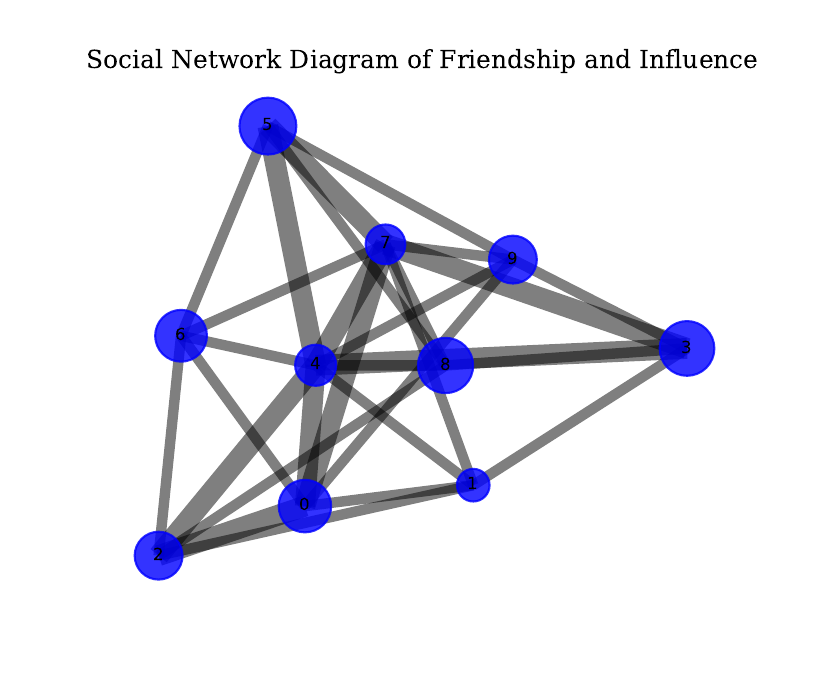}
  \caption{Influence Network}
  \label{fig:sub2}
\end{subfigure}
\caption{Social Network Diagrams representing the relationships and influence dynamics among agents. Figure~\ref{fig:sub1} maps the friendship connections, while Figure~\ref{fig:sub2} details the influence network, with node sizes representing agents' influence levels.}
\label{fig:test}
\end{figure}

Autonomous agents engaged in disaster response simulations rely heavily on the underlying network of relationships and influences. Figures~\ref{fig:sub1} and \ref{fig:sub2} present the complex social dynamics among these agents. Specifically, Figure~\ref{fig:sub1} illustrates the friendship network with nodes and edges representing agents and their friendships, respectively. The diverse colors of the nodes highlight the variety and interconnectivity within this network.

Extending this analysis, Figure~\ref{fig:sub2} delves into the influence dynamics at play. Node size in this figure is proportional to an agent's reputation, indicative of their level of influence. Edges have varying thickness, denoting the strength of trust relationships, with thicker edges signifying higher mutual trust. Collectively, these visualizations elucidate how agents form alliances and the subsequent impact of these relationships on their collective decision-making processes, which are critical for coordinated efforts in disaster response.

\begin{figure}[htbp]
\centering
\includegraphics[width=\linewidth]{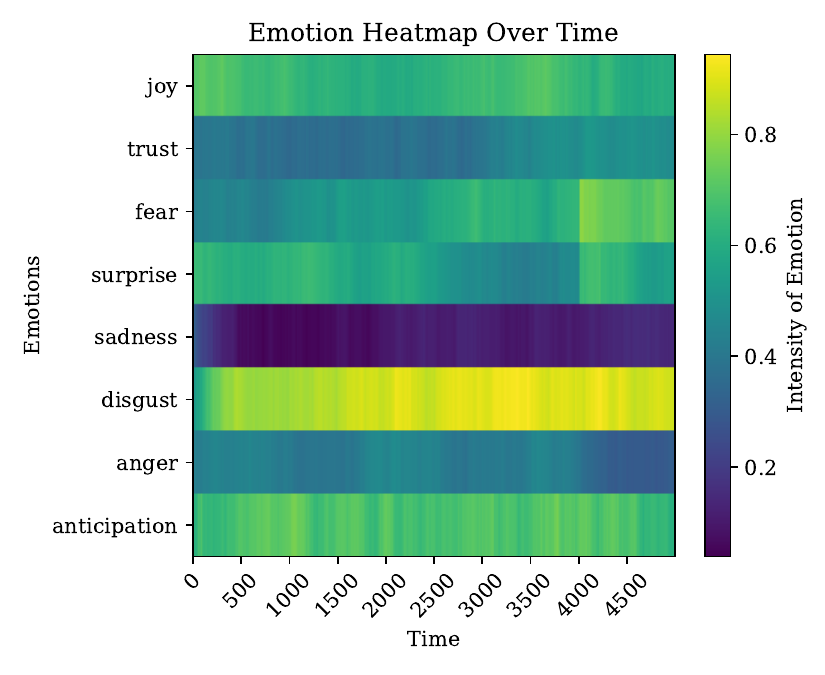}
\caption{Heatmap of various emotions experienced by agents over time, showcasing the intensity and temporal patterns of emotional responses.}
\label{fig:emotionheatmap}
\end{figure}

The heatmap in Figure~\ref{fig:emotionheatmap} captures the intensity and evolution of various emotions experienced by autonomous agents over time within a disaster response simulation. Each row of the heatmap corresponds to a different emotion, with the color gradient reflecting the intensity level of that emotion at various time steps. This visualization offers a nuanced perspective on the agents' emotional fluctuations, shedding light on patterns that might emerge in response to the evolving challenges and circumstances of the simulation.

\begin{figure}[htbp]
\centering
\includegraphics[width=\linewidth]{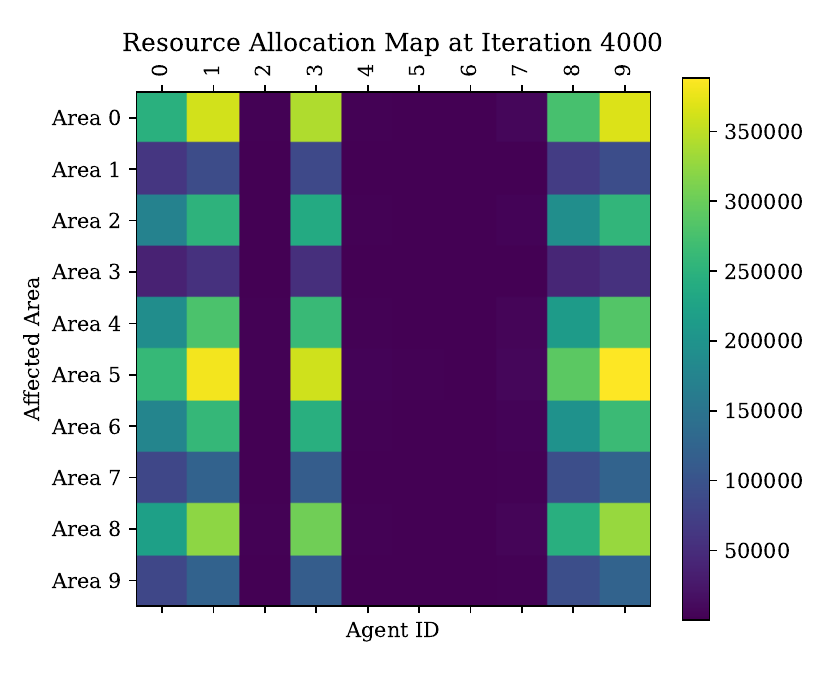}
\caption{Resource Allocation Map at Iteration 4000 showing the distribution of resources to affected areas by individual agents. The color intensity reflects the quantity of allocated resources.}
\label{fig:resourcemap}
\end{figure}

As illustrated in Figure~\ref{fig:resourcemap}, the resource allocation map for iteration 4000 provides a detailed snapshot of how resources were distributed among different areas by each agent. Each column in the map corresponds to an agent, and each row represents an affected area. The intensity of the color indicates the amount of resources allocated, providing an at-a-glance view of the distribution patterns. This visualization is critical for analyzing the efficiency and effectiveness of resource allocation strategies employed by autonomous agents in a simulated disaster response scenario.

\begin{figure}[htbp]
\centering
\includegraphics[width=\linewidth]{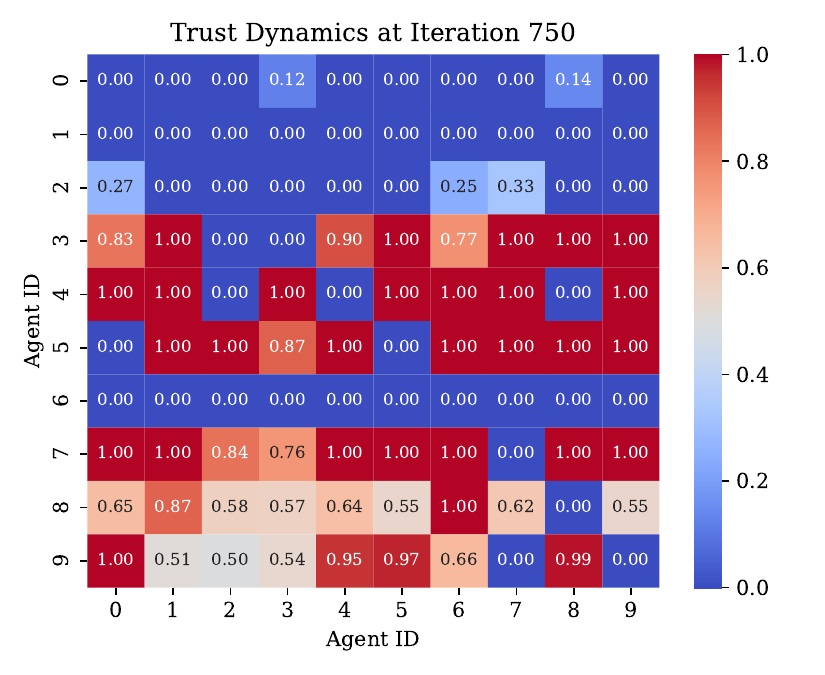}
\caption{Trust Dynamics Heatmap at Iteration 750, illustrating the trust levels between agents. The intensity of the colors indicates the strength of trust, with darker shades representing higher trust levels.}
\label{fig:trustheatmap750}
\end{figure}

During the course of a simulated disaster response, the dynamics of trust among agents is a key factor influencing their interactions and collective decision-making. The heatmap in Figure~\ref{fig:trustheatmap750} provides a detailed view of these dynamics at iteration 750. Each cell in the heatmap corresponds to the trust level from one agent to another, with the color intensity reflecting the strength of trust. This visual representation allows for the identification of patterns in trust building or erosion, which can inform strategies to enhance cooperation among agents.

\section{Discussion}

The Agent-based Cognitive Computational Model, as presented in this paper, stands as a stride in understanding ethical trust in collective decision-making, particularly within disaster response scenarios. Its primary strength lies in the intricate simulation of human-like behavior and interaction, facilitated by Emotional Agents equipped with complex emotional models, memory and learning capabilities, and diverse personality traits. These agents, guided by psychological theories like Plutchik's Wheel of Emotions, offer a nuanced view of how emotions influence ethical decision-making. The integration of social influence, cognitive load, and variable interactions enhances the realism of the simulations, allowing for a detailed examination of trust dynamics within networked environments. The model's ability to incorporate feedback loops and external events further adds to its robustness, providing insights into the non-linear evolution of trust and the impact of unpredictability on decision-making processes. Moreover, the diversity of opinions and strategies employed by agents in the model mirrors the complexity of real-world group dynamics, enabling an exploration of various ethical decision-making approaches. These features make the model a valuable tool for researchers and practitioners in social dynamics, autonomous system design, and the development of ethically aware AI, bridging computational methodology with deep psychological and sociological theory.

Despite its strengths, the model is not without limitations. One notable challenge is the inherent simplification of human trust dynamics in computational simulations. While the model successfully captures specific components of trust, such as trustworthiness, reliability, and reputation, it may not fully encompass the entire spectrum of human emotions and moral complexities. This simplification could potentially lead to gaps in understanding the subtleties of trust development in real-world scenarios. Additionally, the reliance on predefined rules and behaviors for agents, though necessary for simulation, might not account for the unpredictability and variability inherent in human decision-making. Another consideration is the application of the model to real-world scenarios. While the model demonstrates potential in disaster response simulations, its applicability to other contexts may require adjustments and further validation. The model’s dependence on extensive simulations for observing long-term trust evolution also poses challenges in terms of computational resources and time. Lastly, as with any model rooted in theoretical constructs, there is a need for continuous refinement and empirical validation to ensure its relevance and accuracy in depicting complex ethical and trust dynamics.

\section{Conclusion}

In this study, we have ventured through the intricate realms of ethical trust, analyzing its rationality and morality, and culminating in the development of an Agent-based Cognitive Computational Model. This model, integrating advanced emotional frameworks, learning mechanisms, and social dynamics, provides a rich simulation environment to examine the complexities of trust in high-stakes scenarios like disaster response. By embodying cognitive, emotional, and social factors in autonomous agents, we offer insights into the evolving nature of trust and decision-making within socio-technical systems. Our approach not only reinforces the importance of discerning trust but also highlights the potential of computational models in understanding and navigating ethical dilemmas.

Looking ahead, the possibilities for furthering this research are vast. Future work could focus on enhancing the model's realism by incorporating more granular emotional states and complex decision-making processes. There's potential in exploring the application of this model in other high-impact domains such as healthcare, governance, or international relations, where trust dynamics play a crucial role. Another promising direction is the integration of real-world data to refine the model’s predictive capabilities, making it a more potent tool for policy formulation and technology design. Additionally, expanding the model to include more diverse agent interactions and environmental variables could offer deeper insights into the resilience and adaptability of trust networks under various stressors. Finally, ongoing validation and refinement of the model, informed by empirical studies and interdisciplinary collaboration, will be key to ensuring its applicability and relevance in the ever-evolving landscape of ethical trust and decision-making.

\bibliographystyle{ieeetr}
\bibliography{trust}

\end{document}